\documentclass[aps,prl,twocolumn,groupedaddress,showpacs,floatfix]{revtex4-1}%
\usepackage{graphicx,amsmath,amssymb}
\usepackage{amsmath}
\usepackage{amsfonts}
\usepackage{amssymb}
\usepackage{graphicx}%
\providecommand{\U}[1]{\protect\rule{.1in}{.1in}}
\begin{document}
\title{Passivity breaking of a local vacuum state in a quantum Hall system}
\author{Go Yusa}
\email{yusa@m.tohoku.ac.jp}
\affiliation{Department of Physics, Tohoku University, Sendai 980-8578, Japan}
\author{Wataru Izumida}
\affiliation{Department of Physics, Tohoku University, Sendai 980-8578, Japan}
\author{Masahiro Hotta}
\affiliation{Department of Physics, Tohoku University, Sendai 980-8578, Japan}
\date{\today
}

\begin{abstract}
We propose an experimental method for extracting the zero-point energy from a
local vacuum state by performing local operations and classical communication
(LOCC). We model a quantum Hall edge channel as a quantum entangled many-body
channel and the zero-point fluctuation in the charge density of the channel as
the vacuum state. We estimate the order of the energy gain using reasonable
experimental parameters. Such a quantum feedback system breaks the passivity
of a local vacuum state. It can be used to demonstrate Maxwell's demon or
quantum energy teleportation in which no physical entity transports energy.

\end{abstract}

\pacs{03.67.-a, 73.43.-f}
\maketitle




According to quantum mechanics, a many-body system in the vacuum state
possesses a zero-point energy. However, this must be considered a
non-available resource since it is not possible to extract the zero-point
energy from a vacuum by performing local operations ; this is known as the
passivity of the vacuum state \cite{passivity}. However, recent theoretical
studies have suggested that the passivity can be broken \textit{locally} by
performing local measurements followed by local operations and classical
communication (LOCC) \textit{locally}\cite{1}. This scheme can be interpreted
in terms of information thermodynamics as a quantum version of Maxwell's demon
\cite{demon}; specifically, two demons cooperatively extract energy from a
local vacuum state. We consider a quantum entangled many-body system in the
vacuum state. As mentioned above, the demons cannot extract energy from the
system by performing local operations. Here, we define two subsystems A and B
that are separated by an appropriate distance. Since local quantum
fluctuations are entangled in the vacuum state \cite{R}, demon A can obtain a
certain amount of information on quantum fluctuations at B by performing local
measurements at A. However, in exchange for this information, demon A has to
pay an energy $E_{A}$ to his/her own subsystem. Immediately after demon A
informs demon B of the information (i.e., the measurement result), demon B can
extract energy $E_{B}$ even though his/her subsystem remains in a local vacuum
state. This is passivity breaking by LOCC. This scheme is called quantum
energy teleportation (QET) because no actual carriers transfer energy from A
to B but energy can be gained at a remote location B \cite{1}. This type of
quantum feedback is also relevant to black hole entropy, whose origin has
often been discussed in string theory \cite{strominger}, because energy
extraction from a black hole reduces the horizon area (i.e., the entropy of
the black hole \cite{bh}).

To verify this theory in a realistic system requires a dissipationless
quantum-entangled channel with a macroscopic correlation length, a detection
scheme for the vacuum state, and a suitable implementation of LOCC. Quantum
Hall systems are promising systems. They can be formed in a two-dimensional
electron system in a semiconductor subjected to a strong perpendicular
magnetic field. Such a system is suitable because the zero resistance of the
quantum Hall effect means that the system is dissipationless and
quasi-one-dimensional channels appear at the boundary of the bulk
incompressible region of a quantum Hall system; such an edge channel is
considered to behave as a chiral Luttinger liquid \cite{wen}. Power-law
behaviors have been experimentally demonstrated, suggesting that the system
has no certain decay length \cite{chang,grayson}. Furthermore, since the
vacuum is a zero-point fluctuation in the charge density, the Coulomb
interaction (i.e., capacitive coupling), can be used as a sensitive probe for
detecting the vacuum. In addition, semiconductor nanotechnology can be used to
design LOCC.

In this study, we discuss possible passivity breaking in a quantum Hall system
and estimate the order of the energy gain at B by employing reasonable
experimental parameters.\begin{figure}[ptb]
\includegraphics[bb=18 31 208 138, clip,width=8.6cm]{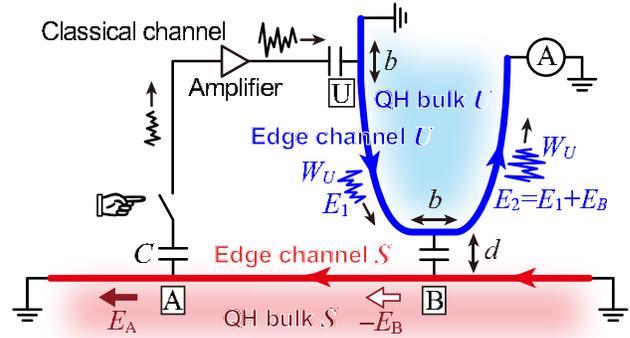} \caption{(color
online). Schematic diagram of the quantum Hall system used in this study. Edge
channels $S$ and $U$ are formed at the boundaries of separate quantum Hall
bulk regions $S$ and $U$, respectively. Red and blue arrows indicate the
directions of propagating waves. $W_{U}$ is a wavepacket excited in region U.}%
\label{fig:model}%
\end{figure}

We consider the system depicted schematically in Fig.~\ref{fig:model}.
Experiments should be performed at low temperatures of the order of
millikelvins (mK) to produce these vacuum states in the edge channels. Two
regions, A and B, are defined on the left-going edge channel $S$. We describe
$S$ by a chiral Luttinger boson field with the trajectory parameterized by a
spatial coordinate $x$ along the channel. Here, we consider the left-going
chiral field whose electron number density $\varrho_{S}(x+v_{g}t)$ deviates
from that of the vacuum state equilibrium. $v_{g}$ is the group velocity of a
charge density wave propagating along the channel (i.e., a magnetoplasmon
\cite{allen} in the zero-energy limit) and $t$ is the elapsed time. The
zero-point fluctuation in the charge density in region A can be experimentally
measured by an RC circuit that consists of the input resistance $R$ of an
amplifier and the capacitance $C$ between $S$ and a local metal gate
fabricated on $S$. For positive operator valued measure (POVM) measurements
\cite{chang}, the RC circuit (i.e., the detector) can be switched on only
during the measurement. $t=0$ is defined as the time when the switch is turned
on. During the measurement, the detector excites $S$ and injects energy
$E_{A}$ into $S$. The measured voltage signal $\upsilon$ is amplified and then
communicated to region U to excite a wave packet $W_{U}$ on the other edge
channel $U$. We describe the trajectory of $U$ by a right-going chiral boson
field $\varrho_{U}(y-v_{g}t)$ parameterized by a spatial coordinate $y$ with
identical velocity $v_{g}$, which assumes that $S$ and $U$ are formed in the
same manner. Thus, $W_{U}$ travels along $U$ carrying energy $E_{1}$ toward
region $B$, where the two edge channels $S$ and $U$ approach each other. These
channels are capacitively coupled only at $B$, where $W_{U}$ interacts with
the zero-point fluctuation of $S$. After the interaction, the energy carried
by $W_{U}$ changes from $E_{1}$\ to $E_{2}$. If no information about
$\upsilon$ is communicated (i.e., $W_{U}$ is created \textit{independently} of
signal $\upsilon$) $W_{U}$ will inject energy to $S$ due to the passivity of
the vacuum state \cite{passivity}. Thus, $E_{B}=E_{2}-E_{1}$ will be negative.
However, in our system, since $W_{U}$ explicitly depends on $\upsilon$, the
passivity is broken and $E_{B}$ is positive; $W_{U}$ gains positive energy
from the zero-point fluctuation of $S$. In the following, we prove this fact
theoretically and estimate $E_{B}$ by setting the experimental parameters
$v_{g}\sim10^{6}$~m/s \cite{ashoori,kamata}, $R\sim10~$k$\Omega$
\cite{impedance}, and $C\sim10$~fF. The length $b$ of regions U and B and the
length of region A are approximated by a typical length scale of $l\sim10~\mu$m.

We start the detailed discussion by modeling the edge channel $S$. The chiral
field operators $\varrho_{S}(x)$ satisfy the commutation relation $\left[
\varrho_{S}(x),~\varrho_{S}(x^{\prime})\right]  =i\frac{\nu}{2\pi}\partial
_{x}\delta(x-x^{\prime})$. The energy density operator of $\varrho_{S}(x)$ is
written as%
\[
\varepsilon_{S}(x)=\frac{\pi\hbar v_{g}}{\nu_{S}}:\varrho_{S}\left(  x\right)
^{2}:,
\]
where $\nu_{S}$ is the Landau level filling factor of $S$ and $::$ denotes
normal ordering, which causes the expectation value of $\varepsilon_{S}(x)$ to
vanish for the vacuum state $|0_{S}\rangle$; $\langle0_{S}|\varepsilon
_{S}(x)|0_{S}\rangle=0$. The free Hamiltonian of $S$ is given by $H_{S}%
=\int_{-\infty}^{\infty}\varepsilon_{S}(x)dx$. The eigenvalue of
$|0_{S}\rangle$ vanishes, $H_{S}|0_{S}\rangle=0$. Taking region A for
$x\in\left[  a_{-},~a_{+}\right]  $ we adopt the RC-circuit-detector model
proposed by F\`{e}ve \textit{et al}. \cite{Feve} to measure the voltage
induced by the zero-point fluctuation of $S$. The charge fluctuation at A is
estimated as
\begin{equation}
Q_{S}(t)=e\int_{-\infty}^{\infty}\varrho_{S}\left(  x+v_{g}t\right)
w_{A}(x)dx, \label{v}%
\end{equation}
with a window function $w_{A}(x)$, which we assume to be a Gaussian function
with its maximum value and width being of the order of unity and $l$,
respectively. In this model \cite{Feve}, the voltage between the detector
contact and $S$ is given by $V(t)=\frac{1}{C}\left[  Q_{S}(t)-Q(t)\right]  $,
where $Q(t)$ is the charge on the capacitor. The quantum noise of the voltage
$V(t=-0)$ is described by the operator $\hat{V}$%
\begin{align*}
\hat{V}  &  =-\sqrt{\frac{\hbar}{\pi RC^{2}}}\\
&  \times\int_{0}^{\infty}d\omega\left[  \frac{\sqrt{\omega}}{\omega-\frac
{1}{iRC}}a_{in}(\omega)+\frac{\sqrt{\omega}}{\omega+\frac{1}{iRC}}a_{in}%
^{\dag}(\omega)\right]  ,
\end{align*}
where $a_{in}(\omega)$($a_{in}(\omega)^{\dag}$) is the annihilation (creation)
operator of excitation of a charge density wave in the detector circuit
satisfying $\left[  a_{in}(\omega),~a_{in}(\omega^{\prime})^{\dag}\right]
=\delta\left(  \omega-\omega^{\prime}\right)  $. Before the measurement (i.e.,
the signal input from $S$ to the detector), $V(t=-0)$ equals $\hat{V}$. Using
the fast detector condition ($RC\ll l/v_{g}$), the voltage after the
measurement is computed as
\begin{equation}
V(t=+0)=\hat{V}+R\dot{Q}_{S}(0), \label{1}%
\end{equation}
where $R\dot{Q}_{S}(0)$ denotes the voltage shift induced by the signal and
the dot of $\dot{Q}_{S}(0)$ indicates the time derivative. The amplitude
$\Delta V$\ of $\hat{V}$ in the vacuum state $|0_{RC}\rangle$ of the RC
circuit can be estimated as $\Delta V=\sqrt{\langle0_{RC}|\hat{V}^{2}%
|0_{RC}\rangle}=O\left(  \sqrt{\frac{\hbar}{RC^{2}}}\right)  \sim10$~$\mu$V.
The root mean square of the voltage shift, $\sqrt{\langle0_{S}|\left(
R\dot{Q}_{S}(0)\right)  ^{2}|0_{S}\rangle}$, is estimated to be $O(100~\mu
$V$)$ from Eq. (\ref{v}), so quantum fluctuations of the edge current will be detectable.

To estimate the energy injected into A after the measurement, we reduce the
measurement operators \cite{chang} to the pointer basis of von Neumann
\cite{vN}. By regarding $\hat{V}$ as a preamplified quantum pointer operator,
the instantaneous shift in Eq. (\ref{1}) can be reproduced by the measurement
Hamiltonian, $H_{m}(t)=\delta(t)R\dot{Q}_{S}(0)P_{\hat{V}}$, where $P_{\hat
{V}}$ is the conjugate momentum operator of $\hat{V}$. Using the eigenvalue
$\upsilon$ of $\hat{V}$ ($\hat{V}|\upsilon\rangle=\upsilon|\upsilon\rangle$),
we can assume the initial wavefunction of the quantum pointer in the
$\upsilon$ representation to be $\Psi_{i}(\upsilon)\varpropto\exp\left[
-\frac{1}{4\Delta V^{2}}\upsilon^{2}\right]  $, whereas the wavefunction after
the measurement is translated as $\Psi_{f}(\upsilon)\varpropto\exp\left[
-\frac{1}{4\Delta V^{2}}\left(  \upsilon-R\dot{Q}_{S}(0)\right)  ^{2}\right]
$. After turning the measurement interaction on, we perform a projective
measurement of $\hat{V}$ to obtain an eigenvalue $\upsilon$ of $\hat{V}$. This
reduction analysis proves that the measurement operator $M_{\upsilon}$
\cite{nc} is $\Psi_{f}(\upsilon)$. The corresponding POVM is represented by
the operator $\Pi_{\upsilon}=M_{\upsilon}^{\dag}M_{\upsilon}$, which satisfies
the standard sum rule, $\int_{-\infty}^{\infty}\Pi_{\upsilon}d\upsilon=I_{S}$,
where $I_{S}$ is the identity operator of the Hilbert space of $S$. The
probability density of the result $\upsilon$ is $p(\upsilon)=\langle0_{S}%
|\Pi_{\upsilon}|0_{S}\rangle$. The post-measurement state of $S$ corresponding
to the result $\upsilon$ is computed to be $M_{\upsilon}|0_{S}\rangle$ up to a
normalization constant. Hence, the average state of $S$ right after the
measurement is given by
\[
\rho_{1}=\int_{-\infty}^{\infty}M_{\upsilon}|0_{S}\rangle\langle
0_{S}|M_{\upsilon}^{\dag}d\upsilon.
\]
The injected energy $E_{A}$ is calculated to be%
\begin{align*}
E_{A}  &  =\int_{-\infty}^{\infty}\langle0_{S}|M_{\upsilon}^{\dag}%
H_{S}M_{\upsilon}|0_{S}\rangle d\upsilon\\
&  =\frac{\hbar v_{g}\nu_{S}}{4\pi}\left(  \frac{ev_{g}R}{2\Delta V}\right)
^{2}\int_{-\infty}^{\infty}dx\left(  \partial_{x}^{2}w_{A}(x)\right)  ^{2}.
\end{align*}
Using the experimental parameters mentioned above, $E_{A}$ is estimated to be
of the order of $1~$meV for $\nu_{S}\sim3$.

We now consider the edge channel $U$. The free Hamiltonian of $U$ is
$H_{U}=\frac{\pi\hbar v}{\nu_{U}}\int_{-\infty}^{\infty}:\varrho_{U}\left(
y\right)  ^{2}:dy$. The measurement result $\upsilon$ is amplified and
transferred to region U as an electrical signal and it then generates a wave
packet $W_{U}$ of $\varrho_{U}(y)$ (i.e., a localized right-going coherent
state) in\ a region with $y\in\left[  b_{-}-L,~b_{+}-L\right]  $. This process
can be expressed by the $\upsilon$-dependent unitary operation $U_{\upsilon}$
to the vacuum state $|0_{U}\rangle$ of $\varrho_{U}(y)$; $|\upsilon_{U}%
\rangle=U_{\upsilon}|0_{U}\rangle$. $U_{\upsilon}$ is given by%

\[
U_{\upsilon}=\exp\left(  \frac{\pi i\upsilon}{\nu_{U}\Delta V}\int_{b_{-}%
-L}^{b_{+}-L}\lambda_{B}(y)\varrho_{U}\left(  y\right)  dy\right)  ,
\]
where $L$ is the distance between regions U and B and $\nu_{U}$ is the filling
factor of $U$. Here, the length $b_{+}-b_{-}=b$ is approximated by $l$. We
assume that $\lambda_{B}(y)$ is a Gaussian function whose maximum,
$\lambda_{B}(b/2-L)$, gives the order of the total number of excess and
deficient electrons from the vacuum-state equilibrium in $\left[
b_{-}-L,~b_{+}-L\right]  $, which are excited by the amplified voltage. We
take the amplitude of $\lambda_{B}\sim O(10)$, which can be experimentally
done by tuning the amplifier gain appropriately. By using $\left[  \varrho
_{U}(y),~\varrho_{U}(y^{\prime})\right]  =-i\frac{\nu_{U}}{2\pi}\partial
_{y}\delta(y-y^{\prime})$, the wave form of $W_{U}$ is computed as
$\langle\upsilon_{U}|\varrho_{U}(y)|\upsilon_{U}\rangle=\upsilon\partial
_{y}\lambda_{B}(y)$. By defining $T$ as the elapsed time when $W_{U}$ has been
generated, the composite state of $S$ and $U$ at $T$ is calculated as
\[
\rho_{SU}=\int_{-\infty}^{\infty}d\upsilon U_{S}M_{\upsilon}|0_{S}%
\rangle\langle0_{S}|M_{\upsilon}^{\dag}U_{S}^{\dag}\otimes|\upsilon_{U}%
\rangle\langle\upsilon_{U}|,
\]
where $U_{S}(T)=\exp(-\frac{iT}{\hbar}H_{S})$. This state is the scattering
input state for the Coulomb interaction between $S$ and $U$. Then, $W_{U}$
evolves into region B by $H_{U}$. The average energy of $W_{U}$ is denoted by
$E_{1}$($=\operatorname*{Tr}\left[  H_{B}\rho_{SU}\right]  $), which is
calculated as%
\[
E_{1}=\frac{\pi\hbar v_{g}}{\nu_{U}}\int_{-\infty}^{\infty}\left(
\partial_{y}\lambda_{B}(y)\right)  ^{2}dy\left[  \langle0_{S}|G_{S}^{2}%
|0_{S}\rangle+\frac{1}{4}\right]  ,
\]
where $G_{S}=-\frac{ev_{g}R}{2\Delta V}\int_{-\infty}^{\infty}\varrho
_{S}\left(  x\right)  \partial_{x}w_{A}(x)dx$. $E_{1}$ is estimated to be of
the order of $10~$meV when $\nu_{S}$ and $\nu_{U}$ are respectively $\sim3$
and $\sim6$. In region B, $W_{U}$ and $\varrho_{S}(x)$ interact with each
other via the Coulomb interaction such that
\[
H_{int}=\frac{e^{2}}{4\pi\epsilon}\int_{b_{-}}^{b_{+}}dx\int_{b_{-}}^{b_{+}%
}dy\varrho_{S}\left(  x\right)  f(x,y)\varrho_{U}\left(  y\right)  .
\]
Here, $\epsilon$ is $10\epsilon_{0}$ for the host semiconductor (e.g., gallium
arsenide), where $\epsilon_{0}$ is the dielectric constant of vacuum. The
function $f(x,y)$ is given by $\frac{1}{\sqrt{\left(  x-y\right)  ^{2}+d^{2}}%
}$. Here, $d$ is the separation length between the two edge channels at B and
is approximated by $l$. After the interaction,\thinspace the energy of $W_{U}$
becomes $E_{2}$. The energy gain, $E_{B}=E_{2}-E_{1}$, is estimated by
lowest-order perturbation theory in terms of $H_{int}$ as follows.%
\begin{align*}
E_{B}  &  =-i\frac{e^{2}v_{g}}{4\epsilon\nu_{S}}\int_{-\infty}^{\infty}%
dz\int_{b_{-}}^{b_{+}}dx_{B}\int_{b_{-}}^{b_{+}}dy_{B}f(x_{B},y_{B})\\
&  \times\int_{-\infty}^{\infty}dt\int_{-\infty}^{\infty}d\upsilon\langle
0_{S}|M_{\upsilon}^{\prime\dag}\varrho_{S}\left(  x_{B}+v_{g}t\right)
M_{\upsilon}^{\prime}|0_{S}\rangle\\
&  \times\langle\upsilon_{U}^{\prime}|\left[  \varrho_{U}\left(  z-v_{g}%
t_{f}\right)  ^{2},~\varrho_{U}\left(  y_{B}-v_{g}t\right)  \right]
|\upsilon_{U}^{\prime}\rangle,
\end{align*}
where $M_{\upsilon}^{\prime}=U_{S}(t_{i}-T)^{\dag}M_{\upsilon}U_{S}(t_{i}-T)$
and $|\upsilon_{U}^{\prime}\rangle=U_{U}(t_{i})^{\dag}|\upsilon_{U}\rangle$.
Here, $U_{U}(T)=\exp(-\frac{iT}{\hbar}H_{U})$. $t_{i}$ and $t_{f}$ are
respectively the start and end times of the interaction between $S$ and $U$.
By substituting the commutation relation given by $\left[  \varrho_{U}\left(
z\right)  ^{2},~\varrho_{U}\left(  y_{B}\right)  \right]  =-i\frac{\nu_{U}%
}{\pi}\partial\delta(z-y_{B})\varrho_{U}\left(  z\right)  $ and integrating
with respect to $z$, we obtain the following relation:%
\begin{align*}
E_{B}  &  =\frac{e^{2}v_{g}}{4\pi\epsilon}\int_{b_{-}}^{b_{+}}dx_{B}%
\int_{b_{-}}^{b_{+}}dy_{B}f(x_{B},y_{B})\\
&  \times\int_{-\infty}^{\infty}dt\partial^{2}\lambda_{B}(y_{B}-v_{g}\left(
t-t_{i}\right)  )\\
&  \times\int_{-\infty}^{\infty}\upsilon\langle0_{S}|M_{\upsilon}^{\prime\dag
}\varrho_{S}\left(  x_{B}+v_{g}t\right)  M_{\upsilon}^{\prime}|0_{S}\rangle
d\upsilon.
\end{align*}
Note that the last integral is computed as

\begin{align*}
&  \int_{-\infty}^{\infty}\upsilon\langle0_{S}|M_{\upsilon}^{\prime\dag
}\varrho_{S}\left(  x_{B}+v_{g}t\right)  M_{\upsilon}^{\prime}|0_{S}\rangle
d\upsilon\\
&  =-\frac{e\upsilon R}{4\Delta V}\int_{-\infty}^{\infty}d\bar{x}_{A}\partial
w_{A}(\bar{x}_{A})\\
&  \times\Delta\left(  \bar{x}_{A}-x_{B}-v_{g}(t+T-t_{i})\right)  + c.c.,
\end{align*}
where
\[
\Delta\left(  x\right)  =\frac{\nu_{S}}{4\pi^{2}}\int_{0}^{\infty}%
dkk\exp\left(  -ikx\right)  .
\]
To integrate $E_{B}$ with respect to $t$, we take the Fourier transform of
$\partial^{2}\lambda_{B}$ in $E_{B}$ and obtain $\partial^{2}\lambda
_{B}(y)=-\frac{1}{2\pi}\int_{-\infty}^{\infty}k^{\prime2}\tilde{\lambda}%
_{B}(k^{\prime})e^{ik^{\prime}y}dk^{\prime}$. Using $\int_{-\infty}^{\infty
}dt\exp[-i(k^{\prime}\pm k)v_{g}t]=\frac{2\pi}{v_{g}}\delta(k^{\prime}\pm k)$,
$E_{B}$ is estimated as
\begin{align}
E_{B}  &  =\frac{3e^{3}vR\nu_{S}}{4\pi^{3}\epsilon\Delta V}\int_{-\infty
}^{\infty}d\bar{x}_{A}\int_{b_{-}}^{b_{+}}d\bar{y}_{B}\int_{b_{-}}^{b_{+}%
}dx_{B}\int_{b_{-}}^{b_{+}}dy_{B}\nonumber\\
&  \times\frac{1}{\sqrt{\left(  x_{B}-y_{B}\right)  ^{2}+d^{2}}}\nonumber\\
&  \times\frac{w_{A}(\bar{x}_{A})\lambda_{B}(\bar{y}_{B}-L)}{\left(
x_{B}+y_{B}-\bar{x}_{A}-\bar{y}_{B}+L+v_{g}T\right)  ^{5}},\nonumber
\end{align}
where $v_{g}T=O(10^{-2}L)$. The parameter $L+v_{g}T(=O(L))$ corresponds to the
distance between A and B. Thus, the energy output $E_{B}$ is estimated to be%
\begin{equation}
E_{B}=O\left(  \frac{e^{2}\lambda_{B}}{4\pi\epsilon l}\frac{ev_{g}R}{l\Delta
V}\left(  \frac{l}{L}\right)  ^{5}\right)  . \label{Eb}%
\end{equation}
Since the function $\lambda_{B}(\bar{y}_{B}+L)$ is positive, $E_{B}$ must also
be positive. Eq. (\ref{Eb}) shows that increasing $L$ will rapidly reduce the
magnitude of $E_{B}$ (e.g., $E_{B}\sim$1$~\mathrm{\mu}$eV for $L\sim4l$).
Nevertheless, for $L\sim2l$, $E_{B}\ $ will be of the order of
$100~\mathrm{\mu}$eV. This is much larger than the thermal energy
$\sim1~\mathrm{\mu}$eV at a temperature of $\sim10~$mK, which is the
temperature at which quantum Hall effect experiments are generally performed
using a dilution refrigerator.

To observe $E_{B}$ experimentally, we measure the current passing through the
edge channel $U$. The relation, $\varepsilon=\frac{\pi\hbar}{\nu_{U}e^{2}%
v_{g}}j^{2}$, between the energy density $\varepsilon$ and the current $j$
gives an energy density of $10$-$\mathrm{\mu eV/\mu m}$, which corresponds to
a current of $10$-$\mathrm{nA}$. This can be detected experimentally using
state-of-the-art electronics. To verify that energy is extracted at B, a
single-shot current measurement should be repeated by switching the circuit on
and off to perform POVM measurements a sufficient number of times to generate
meaningful statistics. In this process, electrical noise, which can be
introduced in the classical channel, is averaged out and thus does not affect
$\langle E_{B}\rangle$.

We now examine energy conservation and dynamics in the system. As we have
shown, extracting $E_{B}$ from the local vacuum state requires measurement
(energy injection) at A. What is the source of $E_{A}$? We consider a POVM
measurement, so that switching the RC circuit injects an energy $E_{A}$into
$S$. Therefore, a battery may provide $E_{A}$to drive the switching device if
the switch is electrically operated. After extracting $E_{B}$, the total
energy $E_{A}-E_{B}$of the system will be non-negative, as it should be
because $E_{A}>E_{B}$. Due to local energy conservation laws, energy transfer
of $E_{B}$from $S$to $U$will result in a negative average quantum energy
density around B. This generation of a negative energy density is attained by
squeezing the amplitude of the zero-point fluctuation less than that of the
vacuum state during the interaction \cite{negativeE}. $-E_{B}$and $E_{A}$will
run "chirally" on the edge toward the downstream electrical ground with
identical velocities of $v_{g}$. Because of this chirality, even after
measurement at region A, $S$around region B will remain in a local vacuum
state with zero energy density.

Unlike quantum Hall systems, several successful experimental studies have been
conducted in quantum optics by introducing LOCC including quantum
teleportation \cite{QT,exp-QT}. Light is a massless electromagnetic
field\ with an infinite correlation length. However, it propagates three
dimensionally so that the energy gain decays rapidly with increasing distance
between A and B. In addition, it is currently difficult to measure the vacuum
state experimentally due to the lack of an appropriate interaction such as the
Coulomb interaction in quantum Hall systems. Thus, our quantum Hall system is
considered to be very suitable for demonstrating local vacuum passivity breaking.

In conclusion, we have theoretically demonstrated that the passivity of the
vacuum can be locally broken by electrical LOCC in a realistic system using a
quantum Hall edge channel as a many-body quantum channel.

\begin{acknowledgments}
The authors gratefully acknowledge K. Akiba and T. Yuge for fruitful
discussions. G. Y., W. I., and M. H. are supported by Grants-in-Aid for
Scientific Research (Nos. 21241024, 22740191, and 21244007, respectively) from
the Ministry of Education, Culture, Sports, Science and Technology (MEXT),
Japan. W. I. and M. H. are partly supported by the Global COE Program of MEXT,
Japan. G. Y. is partly supported by the Sumitomo Foundation.
\end{acknowledgments}


\begin{thebibliography}{99}                                                                                               %


\bibitem {passivity}W. Pusz and S. L. Woronowicz, Commun. Math. Phys.
\textbf{58}, 273 (1978).

\bibitem {1}M. Hotta, Phys. Lett\textit{.} A \textbf{372}, 5671 (2008); M.
Hotta, Phys. Rev. D \textbf{78}, 045006 (2008); M. Hotta, J. Phys. Soc. Jap.
\textbf{78}, 034001 (2009).

\bibitem {demon}W. H. Zurek, in G. T. Moore and M. O. Scully,
\textit{Frontiers of Nonequilibrium Statistical Physics} \ (Plenum Press, New
York), 151, (1984); S. Lloyd, Phys. Rev. A\textbf{ 56}, 3374 (1997); T. Sagawa
and M. Ueda, Phys. Rev. Lett. \textbf{100}, 080403 (2008).

\bibitem {R}B. Reznik, Found. Phys. \textbf{33, }167 (2003); J. Silman and B.
Reznik, Phys. Rev. A\textbf{ 71}, 054301 (2005).

\bibitem {strominger}A. Strominger and C. Vafa, Phys. Lett. B \textbf{379}, 99
(1996); A. Sen, Gen. Rel. Grav. \textbf{40}, 2249 (2008).

\bibitem {bh}M. Hotta, Phys. Rev. D\textbf{ 81}, 044025 (2010).

\bibitem {wen}X. G. Wen, Phys. Rev. B \textbf{43}, 11025 (1991).

\bibitem {grayson}M. Grayson \textit{et al}., Phys. Rev. Lett. \textbf{80},
1062 (1998).

\bibitem {chang}A. M. Chang, L. N. Pfeiffer, and K. W. West, Phys. Rev. Lett.
\textbf{77}, 2538 (1996).

\bibitem {allen}S. J. Allen, Jr., H. L. St\"{o}rmer, and J. C. M. Hwang, Phys.
Rev. B \textbf{28}, 4875 (1983).

\bibitem {ashoori}R.\ C. Ashoori \textit{et al}., Phys. Rev. B \textbf{45},
3894 (1992).

\bibitem {kamata}H. Kamata \textit{et al}., Phys. Rev. B \textbf{81}, 085329 (2010).

\bibitem {impedance}The characteristic impedance of wires is matched to $R$.

\bibitem {Feve}G. F\`{e}ve, P. Degiovanni, and Th. Jolicoeur, Phys. Rev.
B\textbf{ 77}, 035308 (2008).

\bibitem {vN}J. von Neumann, "\textit{Mathematical Foundations of Quantum
Mechanics}", Princeton University Press, (1955).

\bibitem {nc}M. A. Nielsen and I. L. Chuang, "\textit{Quantum Computation and
Quantum Information"}, Cambridge University Press, Cambridge, (2000).

\bibitem {negativeE}Such an emergence of negative energy density has been
widely known in quantum field theory. For example, one of the most simple
cases can be given with linear superposition of the vacuum state and
multi-particle states of a quantum field \cite{Ford}.

\bibitem {QT}C.H. Bennett \textit{et al}., Phys. Rev. Lett. \textbf{70}, 1895 (1993).

\bibitem {exp-QT}D. Bouwmeester \textit{et al}., Nature \textbf{390}, 575
(1997); A. Furusawa \textit{et al}., Science \textbf{282}, 706 (1998).

\bibitem {Ford}L. H. Ford, Proc. R. Soc. London A \textbf{364}, 227 (1978).
\end{thebibliography}
\end{document}